# Impacts of Hot Electron Diffusion, Electron–Phonon Coupling, and Surface Atoms on Metal Surface Dynamics Revealed by Reflection Ultrafast Electron Diffraction

Xing He, Mithun Ghosh, Ding-Shyue Yang*

*Department of Chemistry, University of Houston, Houston, Texas 77204 United States*

*To whom correspondence should be addressed. Email: yang@uh.edu

**Abstract**

Metals exhibit nonequilibrium electron and lattice subsystems at transient times following femtosecond laser excitation. In the past four decades, various optical spectroscopy and time-resolved diffraction methods have been used to study electron–phonon coupling and the effects of underlying dynamical processes. Here, we take advantage of the surface specificity of reflection ultrafast electron diffraction (UED) to examine the structural dynamics of photoexcited metal surfaces, which are apparently slower in recovery than predicted by thermal diffusion from the profile of absorbed energy. Fast diffusion of hot electrons is found to critically reduce surface excitation and affect the temporal dependence of the increased atomic motions on not only the ultrashort but sub-nanosecond times. Whereas the two-temperature model with the accepted physical constants of platinum can reproduce the observed surface lattice dynamics, gold is found to exhibit appreciably larger-than-expected dynamic vibrational amplitudes of surface atoms while keeping the commonly used electron–phonon coupling constant. Such surface behavioral difference at transient times can be understood in the context of the different strengths of binding to surface atoms for the two metals. In addition, with the quantitative agreements between diffraction and theoretical results, we provide convincing evidence that surface structural dynamics can be reliably obtained by reflection UED even in the presence of laser-induced transient electric fields.

## 1. Introduction

Fundamental experimental reports of femtosecond (fs) laser interaction with metals by time-resolved optical methods have been ongoing for about four decades.[1-11] Because of the techniques' high sensitivity to electronic changes, various phenomena were discovered and examined in detail, including a temperature difference between electrons and the lattice,[1,2,12] cooling of the electron temperature to the lattice,[2,3,13] electron–phonon ($e$–ph) coupling strengths for different metals[4,9,14] and their temperature dependency,[5,14] nonequilibrium electron-energy distribution and thermalization,[5-7] as well as an increasing electron-gas energy loss rate with time.[8] Furthermore, the significance of ultrafast transport of nonequilibrium hot electrons as heat carriers has been recognized early.[15-17] These experimental findings generally support the early theoretical consideration of different temperatures for the electron and lattice subsystems in metals at transient times.[18] The widely used two-temperature model (2TM), formulated based on energy balance,[19] has frequently served as a crucial tool for an explanation of laser-induced observations, the extraction of metals' physical constants, or a search for deviations from the presumptions of the phenomenological model. Some further theoretical developments of 2TM were seen to consider the Boltzmann transport equation from a microscopic viewpoint,[20] the inclusion of the ballistic penetration range (for optical absorption) and lattice heat diffusion,[21] temperature-dependent $e$–ph coupling,[22] a nonthermal electronic distribution,[23] the actual density of states in the electronic structure,[24] and nonthermal phonon distributions.[25] However, in the first two decades following the first fs report,[1] most attention was given to the electronic part.

As for the responses of the lattice, surface vibrations of thin metal films as a result of the laser-induced strain and coherent acoustic phonon (CAP) pulse[26] were probed by an optical-beam deflection technique.[27,28] More direct structure-probing measurements using time-resolved diffraction methods were also seen since early years. Initial attempts on metals had a temporal resolution of a few hundred picoseconds (ps),[29] tens of ps,[30,31] to a few.[32,33] The fs regime was later reached to examine CAP dynamics,[34-36] lattice temperature rise and atomic mean-square displacement changes,[25,36,37] and even nonequilibrium phonon dynamics;[25,38] as well as the electronic Grüneisen constant,[39] ultrafast laser melting of metals at the atomic level,[36,40-43] the formation of warm dense matter under high irradiation,[44] the microscopic stress–strain temporal evolution,[45] and most recently, phonon hardening.[46] The majority of these diffraction studies, especially those by ultrafast electron diffraction (UED), were conducted using metal thin films in transmission geometry. The much limited use of UED in reflection geometry (also termed as time-resolved reflection high-energy electron diffraction, tr-RHEED[47]) has provided information about the presence of a weak coupling between bulk and surface vibrations,[48,49] the assessment of phonon confinement effects,[50] as well as phonon scattering dynamics at the nanoscale.[51] Due to the supported bismuth films used in these reports, an interface was taken into account in the discussion of the laser-induced phenomena.

In this report, we use reflection UED for direct structure probing and examine atomic vibrations of metal surfaces following fs laser excitation and the temporal evolution of the corresponding lattice temperatures. Bulk single crystals of gold and platinum with a smooth (111) surface are used. The lack of an interface ensures the consideration of only a single material with its physical constants, avoiding complications from a supporting substrate with different carrier or thermal conductance. It is found that the diffusion of hot electrons is the leading cause for the significant reductions of the surface temperature jump and its decay rate at the surface of gold. While the *e*–ph coupling constant is found to be consistent with the often-used value, appreciably larger *dynamic* vibrational amplitudes of surface atoms are observed for gold, which is reminiscent of similar effects of metal surfaces measured in *thermodynamic* equilibrium.[52-54] In contrast, the surface dynamics of platinum can be well reproduced using 2TM with the metal's accepted physical constants. The main reasons for the behavioral difference between gold and platinum are the former's much lower *e*–ph coupling, higher capability and longer time for hot-electron diffusion, and weaker binding to surface atoms and prominent propensity for deformations.[55] With these experiments, we also note that structural dynamics of materials can be reliably obtained from the data of reflection UED even when a laser-induced transient electric field (TEF) may be present near the sample surface.

2. Methods

Single crystals of Au(111) and Pt(111) with a surface orientation accuracy of <0.1° were obtained from Princeton Scientific. Prior to the dynamics experiments, the crystals were cleaned by several rounds of 10-to-20-minute argon ion sputtering at 1.5 keV at room temperature and 30-minute annealing at about 920 K at a base pressure of the $10^{-8}$ torr level. Afterwards, the crystals were transferred in vacuum without exposure to the ambient. Details about reflection UED have been described.[56-58] Briefly, the ultrahigh vacuum (UHV) chamber assembly had a base pressure of $2\times10^{-10}$ torr. A Yb:KGW regeneratively amplified laser system delivered a fundamental output of 170-fs, 1030-nm pulses at a repetition rate of 10 kHz. The 515-nm (2.41 eV) pulses used to optically excite metals were produced by second harmonic generation (SHG) of the fundamental beam. Another stage of SHG using a fraction of the 515-nm beam produced the ultraviolet (257 nm) pulses, which were focused on a $LaB_6$ emitter tip to generate photoelectron pulses accelerated to 30 keV. The electron diffraction images of the metal surfaces were produced at a grazing incidence angle of $\theta = 1–5°$ and captured by an intensified CMOS camera assembly. The base temperature of the measurements was $T_0 = 290$ K. When needed, the temporal resolution of the reflection UED instrument can reach the sub-ps level with the implementation of a beam-front tilt setup[58] and the use of a reduced number of <100 electrons per pulse. However, more electrons per pulse were typically used to obtain a better signal-to-noise ratio in a short acquisition time. The resulting instrumental response time for the data on ps to nanosecond (ns) scales was about 5.0 ps.

## 3. Results and Discussion

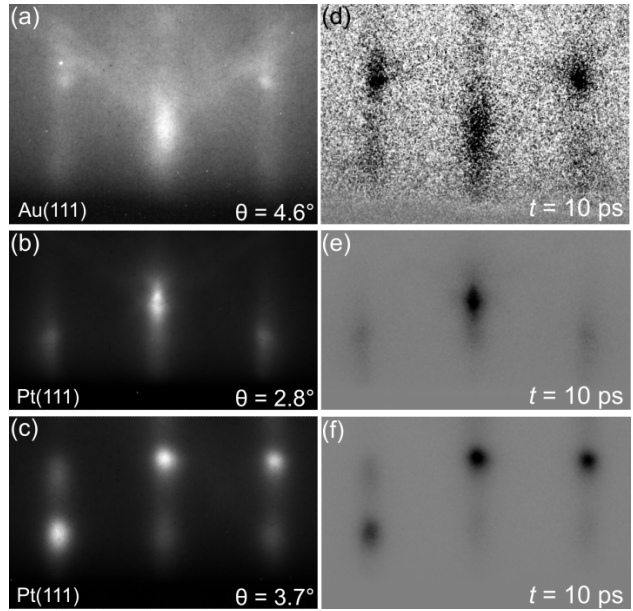

**Figure 1.** RHEED and reflection UED images of sputter–annealed Au(111) and Pt(111) surfaces. (a–c) Steady-state patterns recorded using a gently focused electron beam without laser excitation. Both panel b and c are acquired from Pt(111), with the latter showing a transmission-like pattern due to a rougher surface resulted from omission of the last annealing step after $Ar^+$ sputtering. (d–f) Diffraction differences obtained by subtracting the images recorded at 10 ps by the corresponding negative-time frames in the left panels. Intensity depletion is seen as darkness. Changes near the shadow edge are nearly undiscernible at these indicated incidence angles.

The typical RHEED images of Au(111) and Pt(111) are shown in Figure 1, a and b. The characteristic streaky patterns are the results of smooth Au(111) and Pt(111) surfaces probed by grazing electrons without much penetration. At 30 keV, the elastic mean free paths of electrons are 39.3 and 35.0 Å for Au and Pt, respectively, calculated by $1/N\sigma$ where $N = 4/a^3$ is the number density with $a$ being the lattice constant of the conventional unit cell and $\sigma$ is the element's elastic scattering cross section. The respective probe depths at the (444) Bragg diffraction condition, i.e., $\theta = 3.4°–3.5°$, are 2.33 and 2.13 Å, about the same as the (111) interplanar distances. Thus, compared to the probing over the entire thickness of a thin sample in transmission geometry, RHEED is surface specific and reflection UED is consequently a time-resolved structural method capable of studying light-induced dynamics of the topmost layer or top few layers of a material. In comparison, optical methods such as transient reflectivity and time-domain thermoreflectance have a larger information depth of the order of 10 or more nm.[59,60] Here, annealed Au(111) with the herringbone reconstruction exhibits more evenly distributed diffractions along the streaks (Figure 1a) as a result of rocking of the reciprocal lattice rods due to a spread in the normal vector of the corrugated surface.[61,62] The more concentrated intensity observed from Pt(111) at higher angles is consistent with essentially no surface reconstruction (Figure 1b). However, the still-extended (00) and {11} streaks from the metal and the appearance of additional streaks at low grazing incidence (not shown) are due to an oxygen adlayer chemisorbed during the annealing in a relatively low temperature.[63] To examine potential adlayer influence, we also conduct experiments on sputtered Pt(111) (without the last annealing step), whose transmission-like pattern corresponds to electron penetration through nm-scale surface roughness and allows probing of the near-surface buried layers (Figure 1c).

Shown in Figure 1d–f are the difference images recorded at 10 ps referenced to the respective negative-time frames of Figure 1a–c. The apparent fluences used, $F$, are of the order of 1–10 mJ cm$^{-2}$. Prominent changes are observed for all diffraction features including the RHEED streaks, Bragg spots, Kikuchi lines (if any), and the diffuse scattering background. No fundamental differences in the temporal evolution of the dynamics are found within the fluence range. At sufficiently high $\theta$, changes near the shadow edge are close to an undiscernible level (see later for a discussion about the time-dependent changes due to a laser-induced surface TEF). Thus, the observed UED dynamics must originate from the structural changes following photoexcitation and TEF does not play a noticeable role. To quantify the change at each optical delay time $t$, the diffraction intensity $I(t)$ is extracted by first fitting the vertical profile along a streak or across a spot to determine the center of the peak followed by fitting the horizontal profile across that vertical peak position to a Lorentzian function.

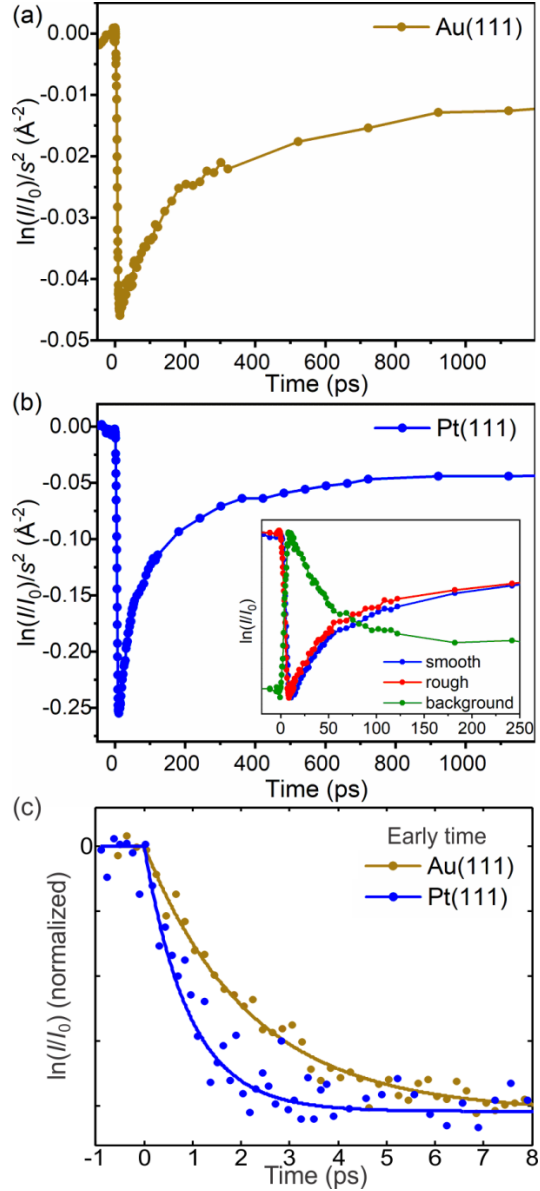

**Figure 2.** Laser-initiated diffraction changes of (a) Au(111) at 10.5 mJ cm$^{-2}$ and (b) Pt(111) at 12.8 mJ cm$^{-2}$. In the inset of (b), the two different surface conditions of Pt(111) give essentially the same dynamics. The diffuse scattering background exhibit the same temporal dependence for the change with an opposite sign. (c) Comparison of the early-time dynamics of Au(111) and Pt(111) measured with <100 electrons per pulse, which indicates the clear difference in the $e$–ph coupling strength. The solid lines are fits of a simple exponential rise function without convolution of the instrument response time.

The ps-to-ns results of $\ln[I(t)/I_0]/s_\perp^2$, where $I_0$ is the original intensity prior to photoexcitation and $s_\perp$ is the vertical momentum transfer of the diffraction, are shown in Figure 2, a and b. It is apparent that Au(111) displays a clearly slower recovery compared to Pt(111), especially in the first 100 ps; the presence of surface roughness does not impact the dynamics of Pt(111) in a fundamental way (Figure 2b, inset). For elemental metals with an isotropic cubic structure and no photoinduced phase transition, the change in diffraction intensity results from the Debye–Waller effect for increased randomized atomic motions,

$$\ln\left[\frac{I(t)}{I_0}\right] = -4\pi^2 s_\perp^2 \cdot \Delta\langle u_\perp^2(t)\rangle \cong -\frac{3h^2 s_\perp^2}{m k_B \Theta_D^2}\Delta T_l(0,t) \qquad (1)$$

where $u_\perp$ is the surface atomic displacement along the surface normal direction, $h$ is the Planck constant, $m$ is the atomic mass, $k_B$ is the Boltzmann constant, $\Theta_D$ is the Debye temperature, and $\Delta T_l(z,t)$ is the lattice temperature increase at depth $z$ and time $t$; $z = 0$ refers to the surface. The model's presumption of thermalized phonons is valid for the temporal range of several ps to ns examined here far beyond the nonthermal period.[25] Moreover, we find that the increase in the diffuse scattering background, which is directly linked to atomic thermal motions, exhibits the same temporal dependence as the decrease of the Bragg diffraction, which provides a crucial justification for the use of the Debye–Waller model (Figure 2b, inset). Thus, the surface temporal evolution of $\Delta T_l(0,t)$ may be derived from the experimental Bragg intensity data, which indicates a nonintuitive result of a slower dissipation for the temperature jump of Au(111) compared to that of Pt(111) even though the thermal conductivity of gold is much higher.

We also record the early-time dynamics of Au(111) and Pt(111) in the first few ps using <100 electrons per pulse for a sub-ps instrumental response time (Figure 2c). A prominent difference is found between the apparent rise times of ~0.9 ps for platinum and 2.2 ps for gold for the initial intensity drop. In what follows we will show that such results are critical to determining the reasons for the anomalies of the Au(111) surface dynamics.

### 3.1 Au(111) surface dynamics

Examination of the influence of hot electron transport is critical to the understanding of the behavior of the Au(111) surface temperature jump and recovery. Without hot electron diffusion, $\Delta T_l(z, t\sim 0)$ would essentially follow the optical extinction profile, i.e. $F(1-R)\alpha/C \cdot \exp(-\alpha z)$, where $R \cong 0.6$ and $\alpha \cong 4.9\times 10^5$ cm$^{-1}$ are the reflectivity and penetration depth of 515-nm light for gold, respectively, and $C = 2.45$ J cm$^{-3}$ K$^{-1}$ is gold's volumetric specific heat (Figure 3a, red curve for the depth dependence). Such an initial temperature profile has been often assumed in various time-resolved studies of materials. However, the above-melting jump of $\Delta T_l = 840$ K with a steep decrease as a function of $z$ would lead to a much faster decay of the surface temperature jump (Figure 3b, red curve), as calculated based on the thermal diffusion equation,

$$\frac{\partial T_l}{\partial t} = D_0 \frac{\partial^2 T_l}{\partial z^2} \tag{2}$$

where $D_0 = \kappa_0/C = 1.29$ cm$^2$ s$^{-1}$ is the thermal diffusivity of gold calculated from the thermal conductivity of $\kappa_0 = 3.17$ W cm$^{-1}$ K$^{-1}$.[9,11] The disagreement with the experimental results is obvious.

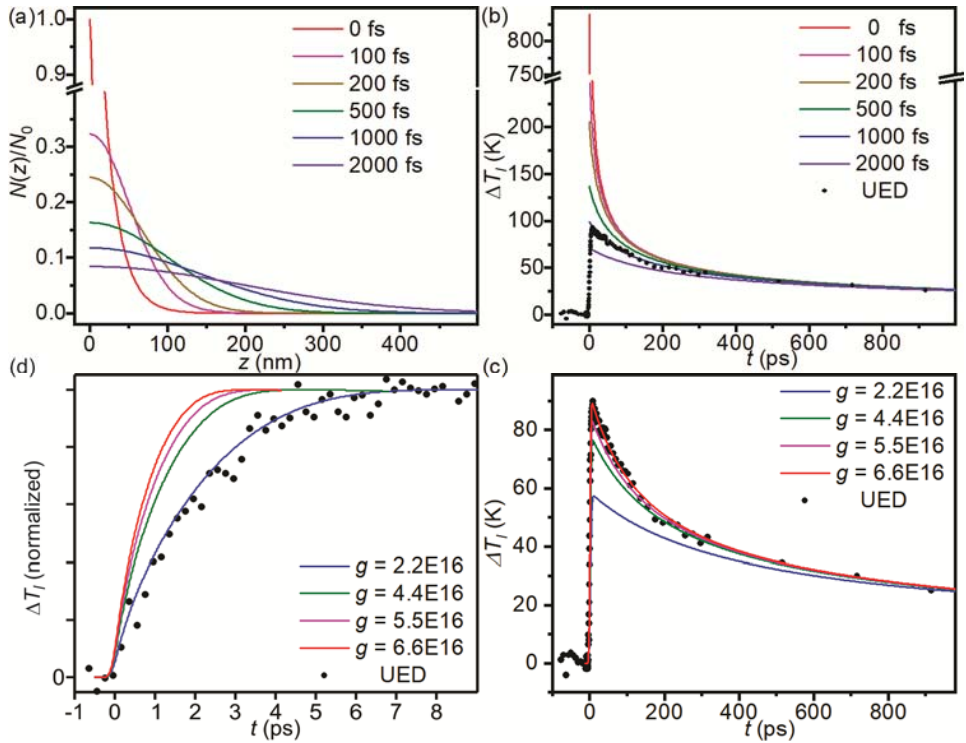

**Figure 3.** Theoretical modeling for the observed temperature jump and evolution of the Au(111) surface. (a) Electron density profiles into the bulk at 10.5 mJ cm$^{-2}$ after diffusion for select durations without energy loss hypothetically. (b) Comparison of the UED-derived surface lattice temperature jump at 10.5 mJ cm$^{-2}$ with the calculated heat diffusion results using the profiles in (a) as different initial conditions. (c) Comparison of the UED-derived surface lattice temperature jump at 10.5 mJ cm$^{-2}$ with the 2TM results calculated using, hypothetically, different $g$ values for e–ph coupling. (d) Comparison of the early-time dynamics with the 2TM results using different $g$ values. The temporal dependence of the data supports the use of the commonly accepted $g$ value.

Now, let us consider an artificial model with two consecutive diffusion stages, first for photoexcited hot electrons, $N(z, t = 0) \propto \exp(-\alpha z)$, to diffuse for a certain duration $t_d$ without losing energy to the lattice, followed by instantaneous conversion of $N(z, t_d)$ into different profiles of $\Delta T_l(z, t_d)$ for the subsequent thermal diffusion described by Equation 2. The benefit of this test model is to assess the importance of hot electron diffusion individually (in a way effectively resembling the case of a photoexcited semiconductor) without the complication by continuous *e*–ph coupling. Using a recently reported effective diffusion constant of 95 cm$^2$ s$^{-1}$,[11] the results of $N(z, t_d)$ with select values of $t_d$ are plotted in Figure 3a. It is clear that the photoinjected energy near the surface is largely reduced even in few hundreds of fs, and the depth profile quickly becomes much broadly distributed. Thus, slower thermal diffusion and hence temperature decay at the Au(111) surface are resulted (Figure 3b). A good match between the UED data and the calculated results with $t_d = 1$ ps signifies the significant extent that hot electron diffusion is required for the observed structural behavior; the initial surface $\Delta T_l \sim 90$ K is merely ~10% of the earlier diffusionless estimate of 840 K and far from the melting point. Moreover, the hot electron dynamics within the first ps imprint the later lattice responses even beyond hundreds of ps. We note that such $t_d$ is a significant fraction of the *e*–ph thermalization period for a few ps and is much longer than the electron characteristic times for, e.g., *e*–*e* scattering and thermalization.[5,23] The long-term effects of the early-time electronic dynamics diminish after a few hundred ps where different $\Delta T_l(z, t_d)$ profiles give nearly the same ns decay trace (Figure 3b).

As a remark, the lattice temperatures shown in Figure 3b are obtained by using Equation 1 on the UED data shown in Figure 2a, followed by a scaling factor to match with the calculated sub-ns results. The need of a non-unity scaler is mostly due to the effect of dynamic scattering by dense, oriented single crystals.[64] However, the confidence in this experiment-theory-corroborated factor is based on the consideration of a known quantity of the injected energy and its conservation and corresponding dissipation in the bulk over time. We also note that further verification comes from the steady-state measurements of the diffraction intensity as a function of temperature to examine the Debye–Waller effect (see Figure S1 and the Supplementary Material). A *static* temperature increase of 85.5 K is obtained for the maximum *dynamical* diffraction intensity decrease seen in Figure 2a, which satisfactorily agrees with and therefore validates our earlier determination of initial $\Delta T_l \sim 90$ K.

Thus, we consider the following form of 2TM as the more realistic theory for Au(111)

that includes e–ph coupling and the diffusion of both hot electrons and the lattice,

$$C_e(T_e)\frac{\partial T_e}{\partial t} = \frac{\partial}{\partial z}\left(\kappa_e(T_e)\frac{\partial}{\partial z}T_e\right) - g(T_e - T_l) + S(z,t) \quad (3)$$

$$C\frac{\partial T_l}{\partial t} = \kappa_l\frac{\partial^2 T_l}{\partial z^2} + g(T_e - T_l) \quad (4)$$

where $C_e(T_e) = \gamma T_e$ is the electronic heat capacity with the electron specific heat constant $\gamma = 71$ J m$^{-3}$ K$^{-1}$ and the temperature of the electron subsystem $T_e$; $\kappa_e(T_e) = \kappa_0 \cdot (T_e/T_l)$ is the temperature-dependent electron thermal conductivity; $\kappa_l = 0.026$ W cm$^{-1}$ K$^{-1}$ is the small lattice contribution of thermal conductivity; $g$ is the e–ph coupling strength; and $S(z,t)$ is the absorbed laser energy density per unit time.[9,11] The one-dimensional nature of the diffusions is due to the experimental condition that the in-plane excitation profile in the $x$ and $y$ directions is largely uniform within the electron-probed region. Shown in Figure 3c are the simulated results with different $g$ values in a hypothetical attempt to find a match, convoluted by a Gaussian function with 5.0 ps as the full-width at half-maximum for the instrumental response time. Intriguingly, the commonly used $g$ value[10] of 2.2×10$^{16}$ W m$^{-3}$ K$^{-1}$ appears to result in a poor match that clearly underestimates the sub-ns recovery rate; the tripled value of $g = 6.6\times10^{16}$ W m$^{-3}$ K$^{-1}$ seems to make a satisfactory agreement in the timeframe of 1 ns. Contradictorily, it is the normal $g$ value that can adequately reproduce the rise time of the initial temperature jump (Figure 3d). This theoretical mismatch between the early- and long-time dynamics needs to be resolved.

We find that a much elevated value of $g$ is not supported by the experimental conditions used in this study; again, the significant role played by electron transport on the Au(111) surface dynamics is noted. A simple energy-based estimate gives $F(1-R)\alpha = \int C_e(T_e)dT_e = \gamma(T_{e,\text{max}}^2 - T_0^2)/2$ when hot electron diffusion is omitted, which would lead to $\Delta T_{e,\text{max}}$ being over 7000 K for $F = 10.5$ mJ cm$^{-2}$ at the surface. However, only about 3400 K is calculated by 2TM when a pulse-width of 200 fs is used, regardless of the $g$ value considered. As a result, the e–ph coupling strength should remain close to the equilibrium value based on the literature, whether or not a nonequilibrium distribution of hot electrons is considered.[24,65-67]

The only way to reconcile the discrepancy in Figure 3c is to consider that the surface temperature jump calculated from the experimental data at early times is larger than predicted by 2TM. It has been understood for decades that a surface does not necessarily behave as a simple termination of the bulk.[53,54] It is well known that the stress-free condition at the surface can lead to increased out-of-plane vibrational amplitudes of topmost atoms compared to those in the bulk.

This means that, effectively, a surface has a lowered $\Theta_D$ than the bulk value, which has been measured by low-energy electron diffraction in thermodynamic equilibrium since 1960s.[52,53] Although reasonably anticipated, the *dynamic* version of this surface behavior has not been directly observed. We note that 2TM disregards any potential differences between the surface and bulk behaviors of a material. Consequently, results of 2TM should be considered as the bulk ones and likely differ from the actual surface lattice dynamics, which is the situation for Au(111). Here, regardless of the scaling factor used (from $\Delta\langle u_\perp^2\rangle$ to $\Delta T_l$) in Equation 1, if one considers the quasi-equilibrium reached between the surface and the bulk at ~1 ns (i.e., the same temperature), the excess $\Delta T_l$ between the UED results and the blue curve in Figure 3c signifes ~50% additional mean-square displacement dynamically exhibited by surface gold atoms initially. A simple exponential decay function can be used to fit the time-dependent difference reasonably well with a time constant of ~125 ps. We also note that other processes such as non-Fourier diffusion (e.g., inclusion of $\partial^2 T_e/\partial t^2$ in Equation 3 to consider wave-like transmission of heat[68]) or ballistic electron transport [e.g., addition of the ballistic range to the optical penetration depth in $S(z,t)$ in Equation 3][21] would further reduce the magnitude and gradient of initial $T_e$, hence resulting in an even lower theoretical $\Delta T_l$ and larger deviation in Figure 3c. This means that our picture of larger dynamic motions of surface atoms is robust even if various aspects of nonequilibrium hot electron dynamics are considered.

We will discuss the possible origin for this Au(111) surface behavior after examination of the Pt(111) surface dynamics.

## 3.2 Pt(111) surface dynamics

The surface structural dynamics of platinum is simpler compared to those of gold that are highly influenced by its high electron conductivity and relatively weak *e*–ph coupling. A good agreement between experimental UED and theoretical 2TM results can already be reached for both high and low fluences using the following accepted physical constants of the metal: $R$ = 91.7% and $\alpha$ = 1.13×10$^6$ cm$^{-1}$ at 515 nm,[69] $\gamma$ = 400 J m$^{-3}$ K$^{-1}$,[37] $\kappa_0$ = 0.70 W cm$^{-1}$ K$^{-1}$, $\kappa_l$ = 0.07 W cm$^{-1}$ K$^{-1}$,[70] $g$ = 3.9×10$^{17}$ W m$^{-3}$ K$^{-1}$,[37] and $C$ = 2.85 J cm$^{-3}$ K$^{-1}$ (Figure 4, a and b). Indeed, the much higher *e*–ph coupling is manifested in the faster lattice temperature jump within the first ps (Figure 4c). However, hot-electron diffusion is still crucial to the explanation of the much reduced surface temperature jumps (less than 95 K instead of 421 K at 12.8 mJ cm$^{-2}$ and less than 10 K instead of 32.9 K at 1.0 mJ cm$^{-2}$) and slowed recovery dynamics not as fast as they would be if the depth profile of optical absorption was considered as the initial condition for thermal diffusion (see the red curve in Figure 3b as an example). In addition, it is again hot-electron diffusion that leads to the paradoxical observation that Pt(111) exhibits a faster apparent recovery than Au(111) at the surface (Figure 2a–b): the spatial gradient of $T_e$ and consequently that of $T_l$ are much reduced for gold compared to platinum because the former's electron thermal conductivity $\kappa_e$ of gold is about 4.5 times the latter's.

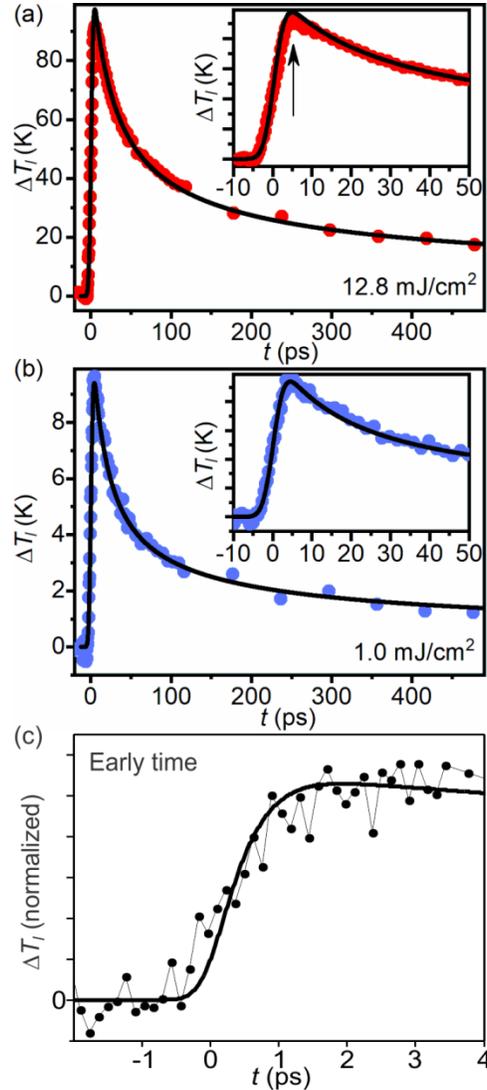

**Figure 4.** Comparison of the observed and calculated temperature jumps and evolution of the Pt(111) surface at (a) 12.8 mJ cm$^{-2}$ and (b) 1.0 mJ cm$^{-2}$ and (c) early time. The temperatures derived from the UED data are shown in dots and the solid curves are the calculated results using 2TM, with convolution with an instrumental response time of 5.0 ps for (a) and (b) and of 0.5 ps for (c). The insets in (a) and (b) show the respective dynamics within the first 50 ps. The arrow in the inset of (a) indicates a small mismatch in the first 15 ps.

Upon closer scrutiny, a small mismatch between the UED and 2TM results is noticed in the first 15 ps for the high-fluence case (arrow in Figure 4a, inset), whereas the agreement is satisfactory for the low-fluence case (Figure 4b, inset). Such a difference may be understood if an additional early-time contribution of the transport of nonthermal hot electrons exists at high fluences, beyond what is taken into account by $\kappa_e(T_e) = \kappa_0 \cdot (T_e/T_l)$. Then the resulting depth profile of $T_e$ will be slightly more spread out into the bulk and, as a result, cause a tiny drop of the maximum $T_l$ and the reduced rate of apparent recovery at short times. However, the influence of this additional factor is not significant given the limited deviation. Thus, we conclude that 2TM with platinum's accepted physical constants is adequate for the surface structural dynamics of Pt(111) in the fluence range used.

It is intriguing that Pt(111) does not exhibit extra dynamic atomic vibrational amplitudes as observed on Au(111). To be clear, Pt(111) surface atoms do have larger out-of-plane motions in equilibrium than those inside the bulk.[53] However, dynamically based on our UED results, the laser-induced transient Pt(111) surface temperature (in association with the surface $\Theta_D$ for the surface atomic motions) appears to be well represented by the bulk value (in association with the bulk $\Theta_D$) calculated by 2TM throughout the observed ps-to-ns temporal window. Hence, the behavioral difference between gold and platinum surface atoms is distinct.

We consider that this may be a good example for strong structure–behavior relationship. The $(22 \times \sqrt{3})$ stripe and larger-range herringbone reconstructions of Au(111) surface have been well reported, with a slight corrugation of 0.2 Å, whereas Pt(111) is unreconstructed at room temperature in vacuum.[71] Recently, Li and Ding reported a theoretical study about the Au(111) herringbone reconstruction and explained the lack of such a structure on Pt(111) as a result of the stronger d–d interaction between Pt atoms and the less stable bridge or top sites to suppress structural rearrangements.[55] It turns out that, dynamically, the surface Pt atoms also behave similarly as the bulk ones. For gold, however, a weaker d–binding between atoms favors the densification of the top layer while limiting the energy penalty owing to the interlayer site mismatches. Incidentally, the local vibrational density of state of the top layer (and only the top layer) is clearly shifted toward lower frequencies along the out-of-plane direction.[55] Such phonon softening strongly suggests the likelihood of larger dynamic vibrational amplitudes of gold surface atoms upon impulsive perturbation, compared to the layers below. Hence, our picture for the observed behavioral difference between the two metal surfaces has a theoretical support.

## 3.3 Surface TEF dynamics

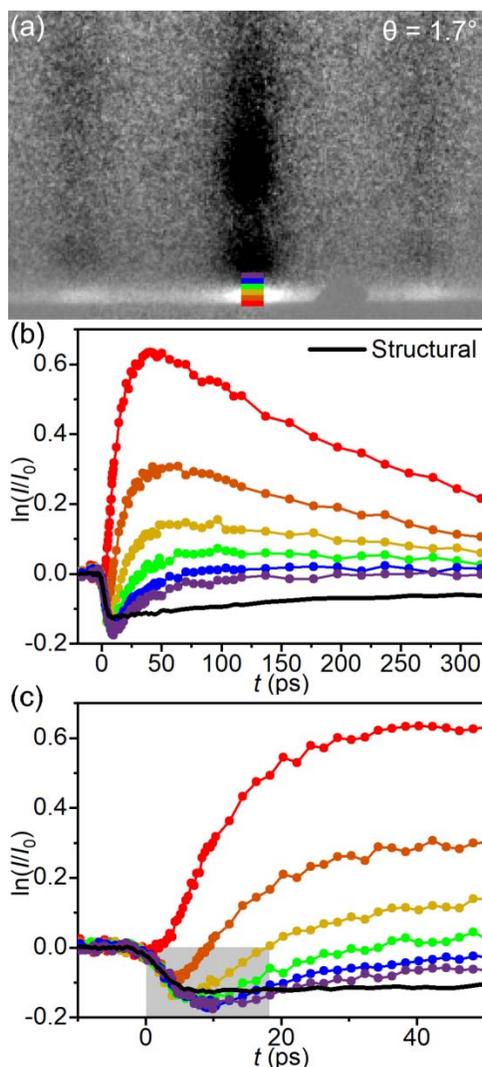

**Figure 5.** Comparison of the diffuse scattering changes near the shadow edge and the structural dynamics. (a) Diffraction image difference of Au(111) at 15 ps at relatively low incidence of 1.7°. Selected near-shadow-edge regions to calculate the relative intensity change as a function of time are denoted by different colors. (b and c) Relative intensity changes near the shadow edge, with the same color codes as in (a), compared to the surface structural dynamics in Figure 2a (black line). The shaded region in (c) for the early-time range highlights the details of the changes within the first 20 ps, which provide crucial clues for the responsible physical processes.

During the developments of UED methodology, Park and Zuo raised the issue of laser-induced TEFs that could affect the trajectories of probing electrons and cast doubt on the interpretation of diffraction changes observed in reflection geometry.[72,73] Since then for years, the time-resolved electron-probe communities regarded reflection UED (tr-RHEED) as problematic and largely stayed away from this mode of detection,[74,75] even though the field-related impacts can be distinguished from actual structural dynamics when care is used.[76,77] In fact, it is worth noting that transmission UED, especially in the non- or low-relativistic regime of the electron speed, is not free from TEF-associated challenges.[78-80] However, the method of ultrafast electron diffractive voltammetry (UEDV) took advantage of such field effects and has been formulated by the Ruan group for more than a decade to probe laser-induced surface photovoltage and charge carrier dynamics, which modulate the scattering angles of surface-penetrating electrons on ultrafast times.[81,82] In recent years, our group has also shown clear evidence that the impact of TEFs is essentially negligible at sufficiently low photoexcitation levels and/or adequately higher incidence angles (leading to diffracted electrons toward higher take-off angles away from the illuminated surface region to minimize the extent and duration of the TEF influence).[56-58,83-85] Given the diffraction geometry, reflection UED provides valuable dynamics information about anisotropic,[58] layered,[57,58,83,84] phase-transition,[56] and molecular thin-film[83,84] materials in the out-of-plane direction that is either difficult to access or challenging to extract from transmission UED data.

In the current study, to once again provide concrete evidence for TEFs' negligible impact on structural dynamics, we compare the aforementioned surface structural dynamics of Au(111) acquired at high $\theta$ with the scattering changes near the shadow edge recorded at a lower $\theta$, where the footprints of the laser excitation and electron probe beams are center-overlapped (Figure 5). The top-layer scattering here is different from the above-surface glancing[72] or displaced overlap[76,86] condition previously used and from UEDV that considers the scattering of a penetrating beam by light-atom materials such as graphite, Si/SiO$_2$, and a self-assembled monolayer of alkylsilanes.[81,82] Evidently, the negligible TEF influence in Figure 1d becomes prominent in Figure 5a due to the accumulated action of more forwardly scattered electrons at grazing incidence by TEFs above the laser-illuminated surface area. Depending on the take-off angles relative to the shadow edge, drastically different temporal dependence of the intensity change can be seen (Figure 5, b and c). Only an increase is found right at the shadow edge because the field of the photoejected electron cloud propagating away from the surface pushes scattered probe electrons toward the surface (Figure 5, b and c, red). Above the shadow edge, the intensity first experiences a sharp decrease due to direct scattering and depletion by the outgoing electron cloud, followed by a gradual increase (similar to the reason above) to different extent with different rise time depending on the distance from the surface (Figure 5, b and c, other colors except red and black). These results are consistent with previous findings with reflection geometry.[72,75,76,79,86] As the TEF influence diminishes, the intensity change significantly reduced at longer times, especially for those regions sufficiently away from the edge (Figure 5b, blue and purple).

However, structural dynamics are entirely different from the TEF-impacted scattering results. First, intensity changes are observed for *all* diffraction-based features with similar temporal behavior soon after photoexcitation regardless of their take-off angles and distances from the shadow edge (Figure 1d–f, and Figure 5a). Second, the remaining changes at hundreds of ps are still prominent, which is inconsistent with the trend of diminishing TEF effects described above (Figure 5b, black and blue). Third and as a crucial clue, the shapes of the dynamics curves at early times indicate the respective physical processes at work (Figure 5c, shaded region). Given that reflection UED probes the surface of a sample and thermal diffusion takes place from the beginning for elemental metals, a recovery of diffraction intensity must occur right after the initial maximum change matures, without any temporal delay or additional growth of intensity drop. This is indeed the case for the Au(111) structural dynamics. In contrast, near-shadow-edge scattering changes exhibit various developments of intensity redistribution due to the finite propagation speed of the outgoing photoejected electron cloud. As a result, a slower, further developing growth of scattering decrease is resulted depending on the relative positions of the two interacting electron groups (Figure 5c, blue and purple in particular in the shaded temporal region). Thus, we believe that with careful choices of experimental conditions and detailed examinations of data, photoinitiated structural dynamics of materials can be reliably obtained by reflection UED even in the presence of non-negligible TEFs.

## 4. Conclusion

In summary, ultrafast electron diffraction (UED) in reflection geometry is used to examine photoinduced structural dynamics of the top layer(s) of Au(111) and Pt(111). The increase and change in the out-of-plane Debye–Waller atomic motions indicate the temporal evolution of the surface lattice temperature, which is quantitatively compared with the theoretical prediction by the two-temperature model that also includes all diffusion components. It is found that even before electron–phonon coupling is dominant to facilitate thermalization, the fast transport of hot electrons greatly reduces surface excitation and consequently affects the recovery dynamics of metal surfaces even till sub-nanosecond times. The two-temperature model is sufficient to reproduce the Pt(111) surface UED results with the accepted physical constants of platinum, which exhibit stronger electron–phonon coupling and moderate electron diffusivity. However, the model is not adequate for the surface dynamics of gold, where the top layer of Au(111) exhibits larger-than-predicted dynamic vibrational amplitudes with a decay time of about 125 ps. The behavioral difference between the two metal surfaces is attributed to their different top-layer structures and interlayer interactions, which demonstrates strong structure–behavior relationship. Finally, from the technical viewpoint, we provide additional evidence to eliminate lingering concerns about the use of reflection UED for structural dynamics as long as care is used in the experimental conditions and data analyses. Given the current understanding of surface dynamics of pristine materials, it will be important to study the impacts of, e.g., adsorbate molecules, adatoms, or molecular overlayers with different binding strengths.

**Supplementary Material**

The supplementary material provides further information about the static measurements of Au(111) diffraction intensity and the Debye–Waller effect.

**Acknowledgments**

The support by the National Science Foundation (CHE-2154363) is acknowledged. The development of the reflection UED instrument at the University of Houston (UH) was partly supported by the R. A. Welch Foundation (E-1860). The theoretical calculations were carried out in part with resources provided by the Research Computing Data Core at UH. One reviewer's suggestion to conduct experiments with sub-ps temporal resolution is acknowledged.

**Conflict of interest**

The authors declare no competing financial interest.

**ORCID**

Xing He https://orcid.org/0000-0001-5341-5662
Ding-Shyue Yang https://orcid.org/0000-0003-2713-9128